\documentclass[aps,twocolumn,noshowkeys,superscriptaddress,noshowpacs]{revtex4}
\usepackage{dcolumn}
\usepackage{amsmath}
\usepackage{bm}
\usepackage{graphicx}
\usepackage{parskip}
\usepackage{epsfig}
\usepackage{color}
\begin{document}

\title{Direct imaging of the ac component of the pumped spin polarization with element specificity}

\author{S. Pile}
\affiliation{Institute of Semiconductor and Solid State Physics, Johannes Kepler University, Altenberger Str. 69, 4040 Linz, Austria}
\author{M. Buchner}
\affiliation{Institute of Semiconductor and Solid State Physics, Johannes Kepler University, Altenberger Str. 69, 4040 Linz, Austria}
\author{V. Ney}
\affiliation{Institute of Semiconductor and Solid State Physics, Johannes Kepler University, Altenberger Str. 69, 4040 Linz, Austria}
\author{T. Schaffers}
\altaffiliation{Present address: NanoSpin, Department of Applied Physics, Aalto University School of Science, P.O. Box 15100, FI-00076 Aalto,
Finland.}
\affiliation{Institute of Semiconductor and Solid State Physics, Johannes Kepler University, Altenberger Str. 69, 4040 Linz, Austria}
\author{K. Lenz}
\affiliation{Helmholtz-Zentrum Dresden-Rossendorf, Institute of Ion Beam Physics and Materials Research, Bautzner Landstr. 400, 01328 Dresden, Germany}
\author{R. Narkowicz}
\affiliation{Helmholtz-Zentrum Dresden-Rossendorf, Institute of Ion Beam Physics and Materials Research, Bautzner Landstr. 400, 01328 Dresden, Germany}
\author{J. Lindner}
\affiliation{Helmholtz-Zentrum Dresden-Rossendorf, Institute of Ion Beam Physics and Materials Research, Bautzner Landstr. 400, 01328 Dresden, Germany}
\author{H. Ohldag}
\altaffiliation{Present addresses: Advanced Light Source, Lawrence Berkeley National Laboratory, Berkeley, CA 94720, USA and
Department of Material Sciences and Engineering, Stanford University, Stanford, CA 94305, USA.}
\affiliation{Stanford Synchrotron Radiation Laboratory, SLAC National Accelerator Laboratory, Menlo Park,
California 94025, USA}
\affiliation{Department of Physics, University of California Santa Cruz, Santa Cruz, California 95064, USA}
\author{A. Ney}
\email[e-mail address:]{andreas.ney@jku.at}
\affiliation{Institute of Semiconductor and Solid State Physics, Johannes Kepler University, Altenberger Str. 69, 4040 Linz, Austria}

\begin{abstract}
Spin pumping in a ferromagnet/nonferromagnet heterostructure is directly imaged with spatial resolution as well as element selectivity. The time-resolved detection in scanning transmission x-ray microscopy allows to directly probe the spatial extent of the ac spin polarization in Co-doped ZnO which is generated by spin pumping from an adjacent permalloy microstrip. Comparing the relative phases of the dynamic magnetization component of the two constituents is possible and found to be antiphase. The correlation between the distribution of the magnetic excitation in the permalloy and the Co-doped ZnO reveals that laterally there is no one-to-one correlation. The observed distribution is rather complex, but integrating over larger areas clearly demonstrates that the spin polarization in the nonferromagnet extends laterally beyond the region of the ferromagnetic microstrip. Therefore the observations are better explained by a local spin pumping efficieny and a lateral propagation of the ac spin polarization in the nonferromagnet over the range of a few micrometers. 

\end{abstract}


\maketitle

In spintronics the generation and manipulation of pure spin currents is in the focus of research activities. Amongst the utilized fundamental effects is spin pumping where a precessing magnetization of a ferromagnet being at ferromagnetic resonance (FMR) transfers angular momentum to an adjacent nonferromagnetic layer \cite{TBB02}. The transfer of angular momentum into the nonferromagnetic layer can be described as a spin-current. The pumped spin current has a dc and an ac component corresponding to the reduction of the projection of the magnetization at FMR as well as the dynamic high-frequency magnetization, respectively \cite{JiB13,WOR14,QRM17}. Usually, spin pumping is electrically detected via the inverse spin Hall effect (ISHE) inside the nonferromagnetic layer which makes conducting high-Z materials such as Pt necessary for detection \cite{ATI11,GMA12,WOR14}; however few other materials like conducting SrRuO$_3$ could be used as well \cite{RPW17}. From the perspective of suitable materials a more versatile approach is to detect the presence of spin pumping via the increased FMR linewidth \cite{UWH01,QRM17,HBM11}. It is nowadays consensus between experiment and theory that the flow of angular momentum from the ferromagnet into the nonferromagnet represents another Gilbert-like damping mechanism. Such spin pumping heterostructures allow different nonferromagnetic materials to be used without the restriction of the ISHE as detection channel. However, the advantages of using nonconducting materials have been pointed out for the ferromagnet \cite{HBM11}, as well as the nonferromagnet \cite{QRM17}. Insulating nonferromagnets have the advantage that the increased magnetic damping is less influenced by other mechanisms like eddy-current damping, see \cite{QRM17} for a recent overview. A major draw-back in general is that most of the experimental methods used so far are detecting spin pumping only indirectly, no matter if they rely on the ISHE or the increased FMR linewidth. In addition, in most approaches only the dc component of the pumped spin polarization inside the nonferromagnet is measured; however also the ac component, which was theoretically predicted to be rather large compared to the dc one \cite{JiB13}, was observed experimentally in a permalloy (Py)/Pt heterostructure via the ISHE \cite{WOR14}; nevertheless studies remain sparse.  

It is the aim of the present work to investigate the ac component of the pumped spin polarization directly inside the nonferromagnet with ultimate spatio-temporal resolution and elemental selectivity. This is achieved by using a time-resolved detection scheme in combination with a scanning transmission x-ray microscope (STXM) \cite{BKC15}. In combination with a microwave excitation this STXM-FMR setup allows to excite the FMR of a ferromagnetic microstructure in contact with an adjacent nonferromagnetic material, which was chosen to be insulating, i.\,e., inaccessible by the ISHE. This combination allows to utilize the known lateral resolution of a few tens of nanometers of the STXM to investigate the lateral extent of the generated spin polarization. This basic approach has already been demonstrated to be feasible for electrical spin injection into a nonferromagnetic metal \cite{KBC15} and very recently also to probe the ac component in Py/Cu heterostructures \cite{DZJ20}. The time-resolved detection scheme based on the internal picosecond time structure of the synchrotron allows to sample dynamics up to the GHz regime, i.\,e. FMR frequencies \cite{BKC15}. Finally, STXM-FMR also allows for probing the magnetic properties with element selectivity, since the contrast mechanism is based on the x-ray magnetic circular dichroism (XMCD) \cite{DEE09}. By that, the dynamic out-of-plane component of the precessing magnetization during FMR can be measured. Therefore, the magnetic response of the driving ferromagnet and the pumped spin polarization inside the nonferromagnetic material can be probed directly and independently.

The system of choice for the present study is a heterostructure consisting of Py in contact with Zn$_{0.5}$Co$_{0.5}$O (50\% Co:ZnO), a system where the presence of spin pumping has been indirectly evidenced before via increased Gilbert damping in FMR \cite{BLN20}. Under the experimental conditions of the STXM-FMR, i.\,e., at ambient temperature, 50\% Co:ZnO is known to be weakly paramagnetic \cite{BLN20} and highly insulating. In addition the $L_3$-edge of Co is in close vicinity in terms of necessary photon energy of the Fe and Ni $L_3$-edges so that both, Py and the Co:ZnO can be conveniently studied with the identical STXM-setup.

Single films of Py, 50\% Co:ZnO, and heterostructures thereof were fabricated on c-plane sapphire substrates as well as SiN-membranes using reactive magnetron sputtering (RMS) at a process pressure of $4\times10^{-3}$~mbar in the same ultrahigh vacuum chamber with a base pressure of $2\times10^{-9}$~mbar. For the 50\% Co:ZnO layer a metallic composite target is used, an Ar:O$_2$ ratio of 10:1 standard cubic centimeters per minute (sccm) is used as process gas, and the substrate temperature is kept at 294$^{\circ}$C, which are optimized growth conditions for 50\% Co:ZnO \cite{BHN19,BLN20}. Py with a typical thickness of 20~nm is fabricated at room temperature using 10 sccm of only Ar as a process gas. To prevent oxidation all Py films are subsequently covered with a 6~nm thick Al capping layer grown under identical conditions using pulsed laser deposition. The feasibility of FMR excitation of a Py microstrip using a micro resonator atop of a Co:ZnO film was tested using a 50\% Co:ZnO film grown on sapphire under identical conditions as above. In a second step a $1 \times 5$~$\mu$m$^2$ Py microstrip with a thickness of 20~nm and an Al cap is fabricated using standard e-beam lithography (EBL). In a final step a planar micro resonator, see \cite{NSS05,NSN08}, with a total Au-plating thickness of 600~nm is fabricated by optical lithography. For the STXM-FMR measurements a $4\times4$~mm$^2$ and 100~nm thick 50\% Co:ZnO film was grown on a commercial 200-nm-thick $0.25 \times 0.25$~mm$^2$ SiN membrane on a highly insulating $5 \times 10$~mm$^2$ Si substrate using shadow masking. Again an identical $1 \times 5$~$\mu$m$^2$ Py microstrip is fabricated using EBL. In a final step a strip-type resonant structure with a total Au-plating thickness of 600~nm is fabricated by optical lithography, see \cite{SMS17} for details. 

\begin{figure}[tb]
\resizebox{1\columnwidth}{!}{\includegraphics{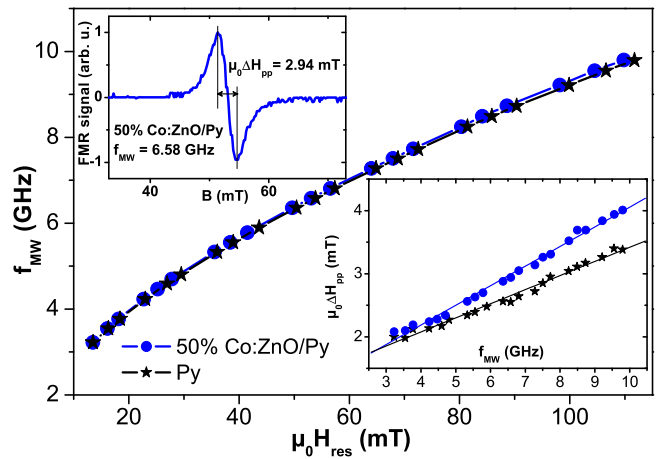}}
\caption{Broadband FMR measurements of the resonance field $\mu_0 H_{res}$ versus microwave frequency $f_{MW}$ for a bare Permalloy (Py) film (stars) and a 50\% Co:ZnO/Py heterostructure both grown on sapphire. The upper inset shows the FMR line and corresponding linewidth $\mu_0 H_{pp}$ at $f_{MW} = 6.58$~GHz. The lower inset summarizes the frequency dependent linewidth. \label{fig1}}
\end{figure}

FMR precharacterization of the chosen materials for the spin pumping heterostructure is done in a home-built broadband FMR setup as reported before \cite{BLN20}. For that an unstructured 50\% Co:ZnO/Py heterostructure and a single Py film grown on c-sapphire substrates are measured using a microwave short adapted after \cite{RMW12} in a frequency range from 3-10~GHz as shown in Fig.\ \ref{fig1}. While the frequency dependence of the resonance field is identical for both samples, the frequency-dependent linewidth exhibits a linear behavior with an increased slope for the heterostructure compared to the Py film. This leads to a Gilbert damping parameter of $\alpha_{Py} = (5.5 \pm 0.3) \cdot 10^{-3}$ for Py and an increased Gilbert damping parameter of $(8.0 \pm 0.3) \cdot 10^{-3}$ for the heterostructure. This has already been discussed before as experimental evidence that angular momentum transfer via spin pumping is possible in Co:ZnO/Py heterostructures \cite{BLN20}. The direct experimental evidence of the presence of a pumped spin polarization in the nonferromagnet is however lacking based on Fig.\ \ref{fig1} only and is inaccessible for the detection via the ISHE.

\begin{figure}[tb]
\resizebox{1\columnwidth}{!}{\includegraphics{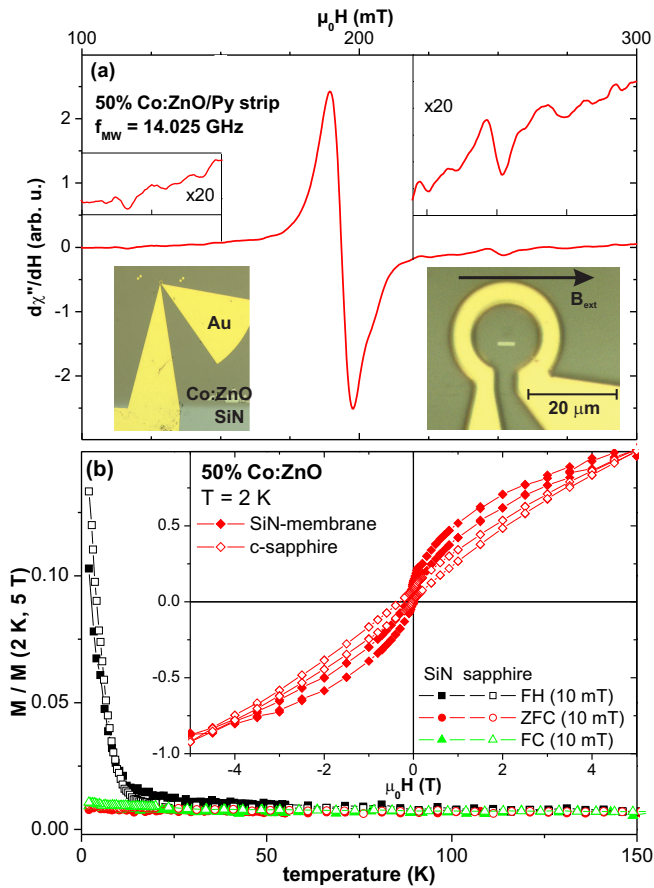}}
\caption{(a) FMR spectrum of the 50\% Co:ZnO/Py strip sample measured at 14.025 GHz; the regions of the spin-vae excitaitons are enlarged. The insets show optical images (overview and close-up) of the planar micro resonator with the strip sample inside. (b) Integral magnetic characterization of 50\% Co:ZnO films grown on sapphire (open symbols) as well as on a SiN-membrane (full symbols). The $M(T)$ curves are shown under different cooling conditions together with the $M(H)$ curves at 2~K in the inset. \label{fig2}}
\end{figure}

To fabricate a spin pumping heterostructure suitable for the STXM-FMR detection scheme, one has to prove in a first step that micro resonator scheme can operate on top of a contingent 50\% Co:ZnO film. For this a Co:ZnO film grown on sapphire with a Py microstrip on top have been studied by a planar micro resonator optimized for 14.025~GHz as shown in the photographs in Fig.\ \ref{fig2}(a). The coventional FMR spectrum in Fig.\ \ref{fig2}(a) exhibits a clear FMR main mode close to 200~mT as well as spinwave exitations above and below the main mode (enlarged by a factor of 20), which have been reported before \cite{BNH11}. The fact that the planar resonator is sensitive enough to detect even these spin wave modes implies that the 50\% Co:ZnO film is sufficiently insulating so that the micro resonator is not shorted out and maintains its high sensitivity.  

In a next step it is verified that the magnetic properties of 50\% Co:ZnO do not significantly change when growing on the SiN-membrane compared to c-sapphire. Figure \ref{fig2}(b) shows superconducting quantum interference device (SQUID) magnetometry measurements following a standard protocol, see \cite{NHL16}. The $M(H)$ curve at 2 K exhibits for both samples the typical, narrowly opened and vertically shifted hysteresis which is indicative of uncompensated antiferromagnetism \cite{NHL16,BHN19}. More important for the STXM-FMR measurements, which are conducted at room temperature only, the $M(T)$ measurements under field heated (FH), field cooled (FC), and zero field cooled (ZFC) conditions reveal that the magnetic order is restricted to below 25~K and no evidence of the formation of Co metallic clusters can be derived by SQUID. It should be noted that for 60\% Co:ZnO grown on SiN the situation is different and a clear superparamagnetic blocking behavior typical for the formation of metallic Co precipitates is seen by SQUID. This is corroborated by conventional STXM measurements around the Co $L_3$-edge, where a heterogeneous appearance of the Co:ZnO film is found for 60\% Co:ZnO, which is absent for the 50\% Co:ZnO (not shown).

\begin{figure}[tb]
\resizebox{1\columnwidth}{!}{\includegraphics{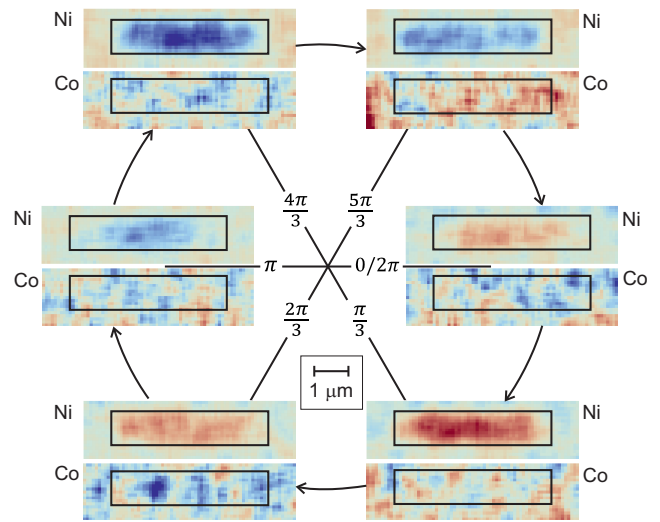}}
\caption{Dynamic magnetic contrast at six relative phases between microwave excitation and x-ray pulse for a 50\% Co:ZnO/Py heterostructure at $f_{MW}=9.447$~GHz and $B_{ext}=96$~mT recorded at the Ni (top images) as well as the Co (bottom images) $L_3$-edges. The region of the Py strip is marked by the black rectangle. \label{fig3}}
\end{figure}

\begin{figure}[tb]
\resizebox{1\columnwidth}{!}{\includegraphics{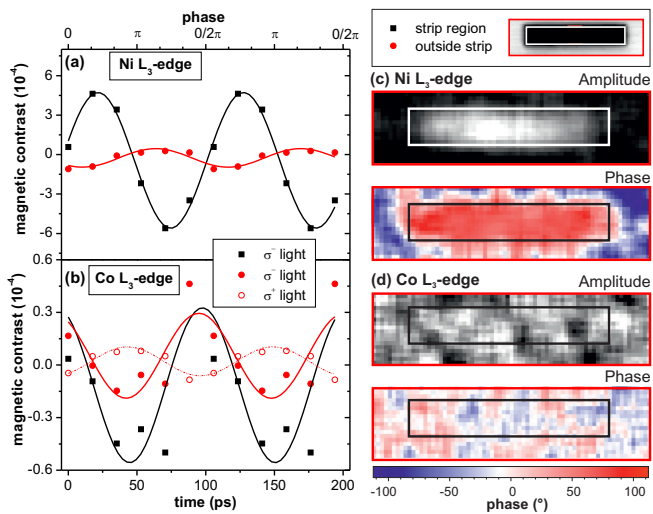}}
\caption{(a) Integrated dynamic magnetic contrast over two prcession cycles recorded for the main FMR mode at $f_{MW}=9.447$~GHz and $B_{ext}=96$~mT recorded at the Ni $L_3$-edge for the region of the strip (black) or outside the strip (red) as marked in the z-contrast image of the Py strip in the inset. (b) Corresponding integrated dynamic magnetic contrast at the Co $L_3$-edge recorded with $\sigma^-$ light; the open symbols correspond to $\sigma^+$ light. The corresponding point-by-point analysis (details see text) of resulting local amplitude and phase derived at the Ni (c) and Co (d) $L_3$-edges. The position of the $1 \times 5$~$\mu$m$^2$ large Py strip is indicated by the white/black frame. \label{fig4}}
\end{figure}

The STXM-FMR measurements themselves have been described in detail in \cite{BKC15,SMS17,SFP19,PFS20}. In brief, the microwave frequency is phase locked to a high harmonic of the internal time structure of the synchrotron to record the x-ray transmission at normal incidence using circular polarized light at six different relative phases between microwave excitation and arrival of the x-ray light pulse. Every other revolution of the electron bunches the microwave power is switched off by a fast PIN-diode so that the six phases are recorded once with precessing magnetization (microwave on) and once with non-precessing magnetization (microwave off). Via the XMCD effect the difference between the images with microwave on and off is proportional to the out-of-plane magnetization component and the six phases thus reflect one full precession cycle of the magnetization in FMR. In Fig.\ \ref{fig3} the dynamic magnetic contrast of a Py microstrip on top of a 50\% Co:ZnO film is shown for a microwave frequency of 9.447~GHz and an external in-plane magnetic field $B_\mathrm{ext}$ of 96~mT, see inset of Fig.\ \ref{fig2}(a). The six consecutive phases were recorded once at the Ni $L_3$-edge (top images) and once at the Co $L_3$-edge (bottom images). The region of the Py strip could be safely derived from the time-integral Z-contrast image at the Ni $L_3$-edge, see white frame in the inset of Fig.\ \ref{fig4}, and is marked by the black frames in Fig.\ \ref{fig3}. It is obvious, that at the Ni $L_3$-edge the quasi-uniform FMR excitation of the Py microstrip can be detected, see \cite{SFP19,PFS20}. Outside the region of the strip little dynamic magnetic contrast is visible. At the Co $L_3$-edge the overall magnetic contrast is much weaker and thus appears more noisy. One can recognize that the Ni signal is strongest at phase $\pi/3$ (mainly red color) and at $4\pi/3$ (mainly blue color) while the Co signal tends to be blueish on the left-hand side and reddish on the right-hand side, i.\,e., the dynamic magnetization of Ni and Co appear to be at antiphase.

For a more in-depth analysis of the observed dynamic magnetic contrast the available magnetic information can be visualized differently. As done before \cite{SMS17,SFP19} the magnetic contrast can be integrated over a region of interest (RoI) and plotted versus (relative) time. For better clarity one full precession period has been duplicated so that two identical consecutive cycles are shown. The result can be seen in Fig.\ \ref{fig4} for the Ni (a) and the Co (b) $L_3$-edge, respectively for two different RoI, the area of the Py strip (black squares) and the region outside the strip (red circles). The data can be fitted with a sine function shown by the lines. At the Ni-edge the dynamic magnetic contrast is nicely visible in the region of the strip. If the region surrounding the strip is averaged only a very small signal is visible. It is important to note that the microwave excitation itself always creates a non-magnetic effect also in the background \cite{SMS17} which is also seen here. This contrast modulation is directly and non-resonantly created by the microwave and, thus, serve as a reference phase of the driving microwave. Now the phase shift is much clearer visible: the quasi-uniform FMR excitation of the Py strip is phase shifted by approximately 90$^{\circ}$ with respect to the microwave which is typical for a driven system close to resonance. 

At the Co $L_3$-edge shown in Fig.\ \ref{fig4}(b) the integrated dynamic magnetic contrast shows significant differences compared to the Ni-edge: (i) the overall amplitude is about a factor of 10 smaller, (ii) the phase at the Co-edge is different from both the background as well as the Ni-edge, and (iii) the contrast is also visible outside the strip region. Observation (i) is understandable since in spin pumping the transfer of the angular momentum across a (non-ideal) interface is presumably suppressed by spin scattering. (ii) demonstrates the important fact that Co has a phase of its own. The phase shift of the Co with regard to the Ni-background, i.\,e., the driving microwave, is about 45$^{\circ}$ which implies that the observed contrast is not a direct effect of the microwave. The fact that Ni an Co are out of phase by close to 180$^{\circ}$ could by understood by adapting the agruments in \cite{TBB02,QRM17} for the conservation of the dynamic part of the angular momentum. This suggests that the pumped spin polarization should be always out of phase with the exciting dynamic magnetic component of the magnetization. (iii) is the most surprising observation. Obviously the dynamic spin polarization is also laterally pumped into the Co:ZnO since it appears to not only exist underneath the Py strip but also outside the region, which is covered by the strip, at least in the measured region (which is only 1 micron larger than the region of the strip). To underline that this small magnetic signal outside the strip region is not an experimental artifact, the circular polarization was flipped from $\sigma^-$ (full symbols) to $\sigma^+$ (open symbols) as a control experiment. The results in Fig.\ \ref{fig4} demonstrate that even outside the strip region the integrated contrast reverses at the Co $L_3$-edge as expected for the XMCD effect. These findings corroborate that the rather noisy dynamic contrast at the Co-edge in Fig.\ \ref{fig3} is indeed a true dynamic magnetic contrast. 

A different way of visualizing the dynamic magnetic contrast was introduced in \cite{GTF19}. For that the time-dependence of the contrast of each pixel is fitted by a sine function and the obtained amplitude and phase of that point-by-point fit is plotted as an image; this is also shown in Fig.\ \ref{fig4} for the Ni (c) and Co (d) $L_3$-edges, respectively. At the Ni $L_3$-edge the amplitude is maximal (white) within the strip with decreasing amplitude towards the edges while it is close to zero (black) outside. The phase has a uniform value across the strip (homogeneously red) corroborating a quasi-uniform excitation characteristic of the main FMR mode in a microstrip, i.\,e., all spins are precessing in phase \cite{BNH11,PFS20}. The situation is different for the Co $L_3$-edge where the overall size of the signal is rather low and, thus, also this visualization appears rather noisy. Nonetheless, it is visible, that while the phase at the Ni-edge is red at the Co it appears bluish, i.\,e., the observation is consistent with the opposite phase visible in the integral dynamic magnetic contrast. It should be noted that all visualizations (except the $\sigma^+$ data) in Figs.\ \ref{fig3} and \ref{fig4} correspond to the identical measurement at the Ni- and at the Co-edge, respectively. Unfortunately, in none of the different visualizations of the data presented so far definite conclusions can be drawn for the actual spatial distribution of the pumped spin polarization inside the Co:ZnO layer. In particular, it is not possible to discriminate between noise and a lateral variation of the spin pumpig efficiency. Only a first evidence of a certain lateral extent of the pumped spin polarization outside the region of the stripe can be provided.  

\begin{figure}[tb]
\resizebox{1\columnwidth}{!}{\includegraphics{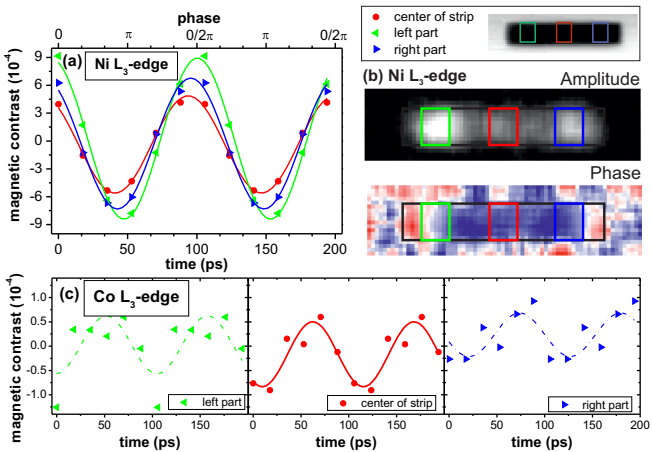}}
\caption{(a) Integrated dynamic contrast for a non-uniform FMR excitation of the Py strip at $f_{MW}=9.447$~GHz and $B_{ext}=106$~mT recorded at the Ni $L_3$-edge for three different regions of interest as marked in the inset. In (b) the corresponding local amplitude and phase is shown. (c) Integrated dynamic magnetic contrast for the three regions of interest recorded at the Co $L_3$-edge. The position of the $1 \times 5$~$\mu$m$^2$ large Py strip is indicated by the black frame. \label{fig5}}
\end{figure}

An alternative approach of using the spatial resolution of STXM-FMR for studying the lateral extent of the pumped spin polarization can be taken by exciting the Py strip non-uniformly. It has been predicted by micromagnetic simulations and verified by the integral FMR measurements \cite{BNH11} that magnetic microstrips can not only be excited quasi-uniformly but also standing spinwaves can exist due to the confined geometry. Their existence for the 50\% Co:ZnO/Py heterostructure has been verified here, see Fig.\ \ref{fig2}(a). These inhomogeneous excitations were already observed in STXM-FMR experiments \cite{SFP19}; however, it turned out that these types of excitation can also have a nonstanding character \cite{PFS20}. In Fig.\ \ref{fig5} the Py strip is excited with the identical microwave frequency of 9.477~GHz but the external field is increased to 106~mT so that a quasi-standing nonuniform excitation is observed. The three different RoI in Fig.\ \ref{fig5} were selected such that they correspond to the maximum amplitude of the point-by-point fitted image as seen in (b). The corresponding phase image reveals only a moderate phase gradient across the strip, especially towards the ends, confirming the quasi-standing character of this particular mode. We first discuss the data recorded at the Ni $L_3$-edge. In Fig.\ \ref{fig5} (a) the integrated intensities of the dynamic magnetic contrast of the three RoI are shown for two consecutive precession cycles. A clear oscillatory behavior is visible with little phase difference for the three regions. Only the left part (green) which exhibits the strongest amplitude is slightly phase-shifted with respect to the two less intense maxima in the amplitude image. This can be interpreted as a slight inhomogeneity of the "proper" resonance condition along the strip. While the left part is nicely driven at resonance the two other regions appear to be slightly off resonance and thus are slightly phase-shifted and reduced in amplitude. This can be caused by a slight variation of the strip thickness, the edge roughness or influences of local stray fields of unknown origin. The fact that local variations of the effective field can alter the behavior of the spinwave modes has already been demonstrated with STXM-FMR \cite{PFS20}. 

For the purpose of this work the nonuniform excitation of the Py strip offers the possibility of investigating the spatial distribution of the pumped ac spin polarization inside the Co:ZnO which is generated by an inhomogeneous distribution of the pumping across the Py strip. Figure \ref{fig5} (c) shows the integrated dynamic spin polarization for the three identical RoI recorded at the Co $L_3$-edge. Interestingly, in both the left (green) and the right (blue) region, where the amplitude of the dynamic magnetization component of the Py is maximum, the signal for the Co is rather noisy and even a proper sine-like behavior is somewhat questionable. However, on the left side (green) obviously phase $0/2\pi$ is a rather bad data point. It should be stressed that all three panels in Fig.\ \ref{fig5} (c) stem from the identical time-resolved experimenal image. A sine fit of these two regions return somewhat reasonable values for phase (roughly opposite to the one of Ni) and amplitude, which point towards a true dynamic magentic contrast. However the quality of the fit is questionable, indicated by the dotted line for the result of the fit. Surprisingly the quality of the fit is significantly improved for the center region (red). Here a rather clear oscillatory behavior of the pumped ac spin polarization can be seen for the Co-edge, which fits a sine function rather satisfingly. It is also out of phase with the one measured at the Ni-edge like for the quasi-uniform mode and thus should represent a true dynamic magnetic contrast generated by spin pumping. It is remarkable, that this clear signal can be seen at the Co $L_3$-edge in regions where the actual amplitude of the pumping itself as measured at the Ni $L_3$-edge is rather weak which again points towards a lateral extent for the pumped ac spin polarization. 

Our findings for the spatial extent of the dynamic spin polarization measured inside the nonferromagnetic material with element specifity provides direct experimental evidence for spin pumping into an insulating oxide. Thus, it corroborates the interpretation that the increased Gilbert damping seen in intergal FMR in Fig.\ \ref{fig1} is indeed a sign of spin pumping. In addition utilizing the spatial resolution of the STXM-FMR it is obvious that no one-to-one correspondence exists with the spatial distribution of the dynamic behavior of the pumping ferromagnetic material. The results can be merely understood in the frame of a local pumping efficiency which generates the inhomogeneous appearance of pumped ac spin polarization. In particular, we find evidence that it can laterally spread out inside the nonferromagnetic material for finite distances as indicated in Fig.\ \ref{fig4}. It even seems to form interference effects as suggested by the unexpectedly large and clear signal in the center of the strip despite the weak pumping in this very region as seen in Fig.\ \ref{fig5}. The lateral extent of the pumed ac spin polarization is different from previous findings of the dc- \cite{KBC15} as well as the ac-component \cite{DZJ20} in Py/Cu heterostructures, where the polarization was confined to the area of the ferromagnet. A substantial lateral extent of the ac component of the pumped spin polarization inside the nonferromagnetic insulator can be supported by another consideration. Unlike in ISHE experiments, a high-Z metallic material like Pt or Pd, can be avoided and we can use Co:ZnO as a highly insulating and rather low-Z material. Therefore the spin relaxation times and/or the spin coherence length can be expected to be much larger than for example in metallic Cu, where the expected decay-length is of the order of 100~nm and is therefore not easy to be detected by the spatial resoluton of the STXM \cite{KBC15,DZJ20}. In fact long spin coherence lengths/times at room temperature are well known for ZnO \cite{GSL05} from optical measurements; obviously the presence of the paramagnetic Co ions does not severely reduce the spin coherence. 


In summary, we demonstrate the feasibility to investigate the lateral distribution of the pumped ac spin polarization inside a Co:ZnO/Py heterostructure with element selectivity. A Py microstrip is driven into ferromagnetic resonance of either the quasi-uniform main excitation or a quasi-standing nonuniform magnetic excitation seen in intergal FMR and directly verified by STXM-FMR. The dynamic spin polarization inside the adjacent nonferromagnet which is generated via spin pumping can be directly imaged with unprecedented spatio-temporal resolution. While the very existence of spin pumping as suggested by integral FMR studies can be unambiguously corroborated, the spatial extent of the spin polarization appears to be more complex and the findings are consistent with a locally varying spin pumping efficiency. More significantly the dynamic spin polarization in the Co:ZnO can be detected outside the region of the Py strip thus inferring a laterally expanding spin polarization inside the nonferromagnet.   

\begin{acknowledgments}
The authors would like to thank the Austrian Science Foundation (FWF), Project No. I-3050, for financial support. Use of the Stanford Synchrotron Radiation Lightsource, SLAC National Accelerator Laboratory, is supported by the U.S. Department of Energy, Office of Science, Office of Basic Energy Sciences under Contract No. DE-AC02-76SF00515. T. Feggeler (Univ. Duisburg-Essen) is gratefully acknowledged for help during the SSRL beamtime.
\end{acknowledgments}

\end{document}